# Nesting-driven antiferromagnetic order in Kondo lattice CePd$_5$Al$_2$

Chen Zhang,[1] Ya-Hua Yuan,[1] Jiao-Jiao Song,[1] Ján Rusz,[2] Yin-Zou Zhao,[1] Qi-Yi Wu,[1] Yu-Xia Duan,[1] Yasmine Sassa,[2,3] Oscar Tjernberg,[4] Martin Månsson,[4] Magnus H. Berntsen,[4] P. H. Tobash,[5] Eric D. Bauer,[5] Peter M. Oppeneer,[2] Tomasz Durakiewicz,[6] and Jian-Qiao Meng[1,*]

[1]*School of Physics and Electronics, Central South University, Changsha 410083, Hunan, China*
[2]*Department of Physics and Astronomy, Uppsala University, Box 516, S-75120 Uppsala, Sweden*
[3]*Department of Physics, Chalmers University of Technology, 41296 Göteborg, Sweden*
[4]*Department of Applied Physics, KTH Royal Institute of Technology, Electrum 229, SE-16440, Stockholm, Kista, Sweden*
[5]*Los Alamos National Laboratory, Los Alamos, New Mexico 87545, USA*
[6]*Idaho National Laboratory, Idaho Falls, ID 83415 USA*
(Dated: Monday 10$^{\text{th}}$ April, 2023)

We investigated the electronic structure of the antiferromagnetic Kondo lattice CePd$_5$Al$_2$ using high-resolution angle-resolved photoemission spectroscopy. The experimentally determined band structure of the conduction electrons is predominated by the Pd 4$d$ character. It contains multiple hole and electron Fermi pockets, in good agreement with density functional theory calculations. The Fermi surface is folded over $\bm{Q}_0$ = (0, 0, 1), manifested by Fermi surface reconstruction and band folding. Our results suggest that Fermi surface nesting drives the formation of antiferromagnetic order in CePd$_5$Al$_2$.

Charge order exists widely in correlated materials and can be observed in real and momentum space. It often alters the low-energy electronic structure near the Fermi energy ($E_F$), leading to the reconstruction of the Fermi surface (FS), which is often the key to discriminating between different theoretical models [1–3]. In a three-dimensional (3D) system, FS nesting gives rise to very rich low-temperature phases due to quasiparticle interaction. FS nesting has been widely studied in high-temperature superconductors [2, 4], charge density waves (CDW) [5–7], and heavy Fermion (HF) compounds [8–11].

In HF compounds, the competition between the Kondo effect and the Ruderman-Kittel-Kasuya-Yosida (RKKY) interactions determines the ground state, giving rise to a rich variety of exotic phenomena such as unconventional superconductivity [13], hidden order [14], CDW [15], quantum criticality [16], topological insulator [17], and others. The RKKY interaction promotes magnetic order phases, such as antiferromagnetism (AFM) or ferromagnetism (FM). FS nesting has been proposed both theoretically [18, 19] and experimentally [8–11, 39] in HF compounds. It is believed that FS nesting can generate itinerant AFM, and mismatching of nesting can suppress the order and tune the Néel temperature $T_N$ to zero, resulting in quantum critical points [18].

The HF compound, CePd$_5$Al$_2$, is an AFM superconductor identified in 2007 [20]. It undergoes two AFM transitions at $T_{N1}$ = 3.9 K (or 4.1 K) and $T_{N2}$ = 2.9 K [20–22]. The crystal structure of CePd$_5$Al$_2$ is a tetragonal ZrNi$_2$Al$_5$-type structure with the space group $I4/mmm$, in which the CePd$_3$ and Pd$_2$Al$_2$ layers are stacked along the $c$-axis. The intermediate Sommerfeld coefficient $\gamma$, 60 mJ/(mol K$^2$) for polycrystals [20] or 18 mJ/(mol K$^2$) for single crystals [23], indicates that CePd$_5$Al$_2$ is not a conventional HF compound. Transport experiments suggested that the Ce 4$f$ electrons are well-localized at low temperatures [23–25]. Below $T_{N1}$, an in-plane modulated incommensurate magnetic structure [$q$ = (0.235, 0.235, 0)] was revealed by single crystal neutron scattering [22, 26]. The magnetic momentums of CePd$_5$Al$_2$ have been suggested to align along the [001] direction with an AFM Ising-like magnetic structure [23, 24], which is unfavorable for superconductivity. However, for magnetically mediated HF superconductivity, quasi-two-dimensional (quasi-2D) compounds are likely to have a higher transition temperature than 3D compounds [27, 28]. Due to the layered structure and large interlayer distance in CePd$_5$Al$_2$, theoretical calculations [24, 25] and angle-resolved photoemission spectroscopy (ARPES) measurements [29] suggested that some FSs have quasi-2D characteristics. Bulk superconductivity of CePd$_5$Al$_2$ was induced by applying pressure in the range of 9−12 GPa, reaching a maximum of $T_c$ = 0.57 K [21]. To better understand the HF physics, superconductivity, and AFM in this material, we examine the FS topology and low-energy electronic structure of CePd$_5$Al$_2$ using ARPES.

In this Letter, we have systematically investigated the electronic structure of the AFM superconductor CePd$_5$Al$_2$ utilizing high-resolution ARPES measurements and density-functional theory (DFT) calculations. Using tunable photon energies of synchrotron radiation, the shapes of the FSs near Γ and Z were mapped out. The measured FS topology shows good agreement with the DFT calculations, indicating the dual nature of the 4$f$ electrons. Constant photon energies ($h\nu$ = 87 and 100 eV) mapping suggested a twofold symmetry of the low-lying electronic structure along the $c$-axis. Our data suggest that the AFM ordering in CePd$_5$Al$_2$ roots in a FS nesting instability.

High-quality single crystals of CePd$_5$Al$_2$ were synthesized by the arc-melting method as described elsewhere [29]. ARPES measurements were performed at the SIS X09LA beamline of the Swiss Light Source using a VG-SCIENTA R4000 hemispherical electron energy analyzer. All samples were cleaved *in situ* along the (001) plane at low temperature and measured in an ultrahigh vacuum with a base pressure better than $4 \times 10^{-11}$ mbar. Photon energies of 87 and 100 eV were chosen to probe the FSs ($k_x$-$k_y$ plane), where $k_z$ is close to the Γ ($k_z \sim 0.2 \times 2\pi/c$) and Z ($k_z \sim 0.9 \times 2\pi/c$) points, respectively, estimated based on an inner potential $V_0$ of 16 eV [29]. All ARPES data were collected at a low temperature of 10 K with an angular resolution of 0.2°. Our data are compared with a DFT band-structure calculation performed in the nonmagnetic state.

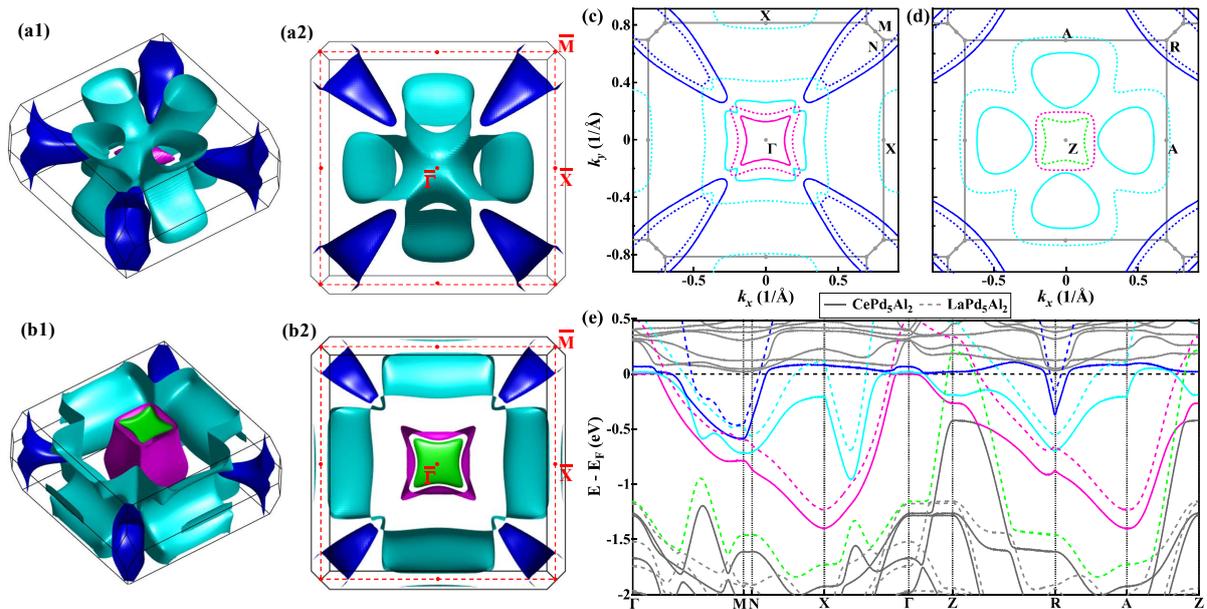

FIG. 1. (color online) Comparison of calculated electronic structure between CePd$_5$Al$_2$ and LaPd$_5$Al$_2$. (**a1**) and (**a2**) side and top views of the 3D bulk FSs of CePd$_5$Al$_2$, respectively. (**b1**) and (**b2**) side and top views of the 3D bulk FSs of LaPd$_5$Al$_2$, respectively. The red-dashed line in (a2) and (b2) depicts the projected surface BZ for (001) surface. Here different colors stand for different FS sheets. (**c**) and (**d**) Calculated 2D FS contours at $k_z = 0$ plane and $k_z = 2\pi/c$ plane, respectively. (**e**) The calculated band structure along high-symmetry direction of the BZ. The $E_F$ is set to 0 eV. For comparison, solid lines and dashed lines represent the bands of CePd$_5$Al$_2$ and LaPd$_5$Al$_2$, respectively. The color of different FSs is consistent with the colored bands that cross the $E_F$.

The calculated 3D FSs of CePd$_5$Al$_2$ in the nonmagnetic phase are shown in Figs. 1(a1) and 1(a2) as side and top views, respectively (See the Supplemental Materials [30]). Three FS sheets of different colors can be seen. The CePd$_5$Al$_2$ FSs topology changes are strong and complicated (see Fig. S1 of the Supplemental Material [30] for more details). The FS sheets in magenta and cyan in the zone center region show obvious 3D characteristics. While the large FS sheet in blue centered at Brillouin zone (BZ) corner M is corrugated but cylindrical in topology. Similarities and differences exist between our computed FSs and previous ones [24, 25]. All calculations propose the cyan and blue FSs. However, the previous calculations do not give the magenta FS around the Γ point but proposed a 4$f$-electron component-dominated cylindrical FS at the BZ corner. For comparison, Figs. 1(b1) and 1(b2) show the calculated results of the hypothetical non-$f$ reference compounds LaPd$_5$Al$_2$, using the lattice parameters of CePd$_5$Al$_2$. Our calculation is in agreement with others [24, 25]. Four FSs are observed in Fig. 1(b). The topology of the LaPd$_5$Al$_2$ FSs shows more 2D characteristics. The FSs of CePd$_5$Al$_2$ have some similarities with those of LaPd$_5$Al$_2$, such as the blue FS in the corner of the BZ. While the magenta cylindrical FS of LaPd$_5$Al$_2$ shrinks to a pillow-shaped FS in CePd$_5$Al$_2$, the green pillow-shaped FS of LaPd$_5$Al$_2$ disappears completely in CePd$_5$Al$_2$, and the cyan FS of LaPd$_5$Al$_2$ shrinks in volume in CePd$_5$Al$_2$.

Figures 1(c) and 1(d) show a comparison of the calculated 2D FS topology for CePd$_5$Al$_2$ and LaPd$_5$Al$_2$ at $k_z = 0$ and $k_z = 2\pi/c$ planes, respectively. The FS topology of CePd$_5$Al$_2$ and LaPd$_5$Al$_2$ shows a significant change with $k_z$. The overall shapes of the FSs of CePd$_5$Al$_2$ and LaPd$_5$Al$_2$ are very similar, but the size difference is huge. Figure 1(e) shows the calculated band structure along several high symmetry directions of the BZ. Multiple bands are present in the 2 eV energy range. Ce 4$f$ orbitals dominate the spectrum just above the Fermi energy ($E_F$), as evidenced by the flat bands. The presence of the Ce 4$f$ state causes a strong modification of the band structure near the $E_F$. As a result, for the low-lying occupied states, the band structure of CePd$_5$Al$_2$ (solid lines) is quite different from that of LaPd$_5$Al$_2$ (dashed lines), i.e. not only due to shifts in energy. It is similar to the case of CeRu$_2$Si$_2$ and LaRu$_2$Si$_2$ [33], but different from the case of CePt$_2$In$_7$ and LaPt$_2$In$_7$ [34, 35] as well as CeIrIn$_5$ and LaIrIn$_5$ [36].

Figure 2 presents the electronic structure measured with a constant photon energy of 87 eV, corresponding to a cut near the X-Γ-X plane ($k_z \sim 0.2 \times 2\pi/c$) [indicated in light blue in Fig. 2(a)]. The measured electronic structure is shown in the 3D volume plot [Fig. 2(b)], which shows the band disper-sions along the high-symmetry directions. Figure 2(c) shows constant-energy contours of the band structure at different binding energies. The FS contours are shown with false color. To facilitate comparison between the measured and calculated FS, the corresponding calculated FS contours are plotted as solid colored lines in Fig. 2(d). It can be seen that the measured and calculated FS contours are in reasonable agreement concerning the shape of the pockets around Γ, indicating the involvement of 4$f$ electrons in the formation of the FS. This implies that the 4$f$ electron exhibits the dual nature of being both itinerant and localized, which is a common feature of Ce-based heavy fermions [35, 37–39]. Multiple FS sheets were revealed. Two hole-like pockets centered on $\overline{\Gamma}$ are observed, which is in good agreement with the band structure calculation. Around the $\overline{M}$ point of the BZ, a "double" pocket can be seen [Fig. 2(c1)], inconsistent with the band structure calculations. The solid blue curve in Fig. 2(d1) shows that the calculations predict an $\overline{M}$-centered rugby-ball-shaped electron pocket with its long axis pointing to the $\overline{\Gamma}$ point. Our results also differ from the previous calculations [24, 25], which show that in addition to the rugby-ball-shaped pocket, there is an $\overline{M}$-centered circular electron pocket composed of a significant component of 4$f$ electrons.

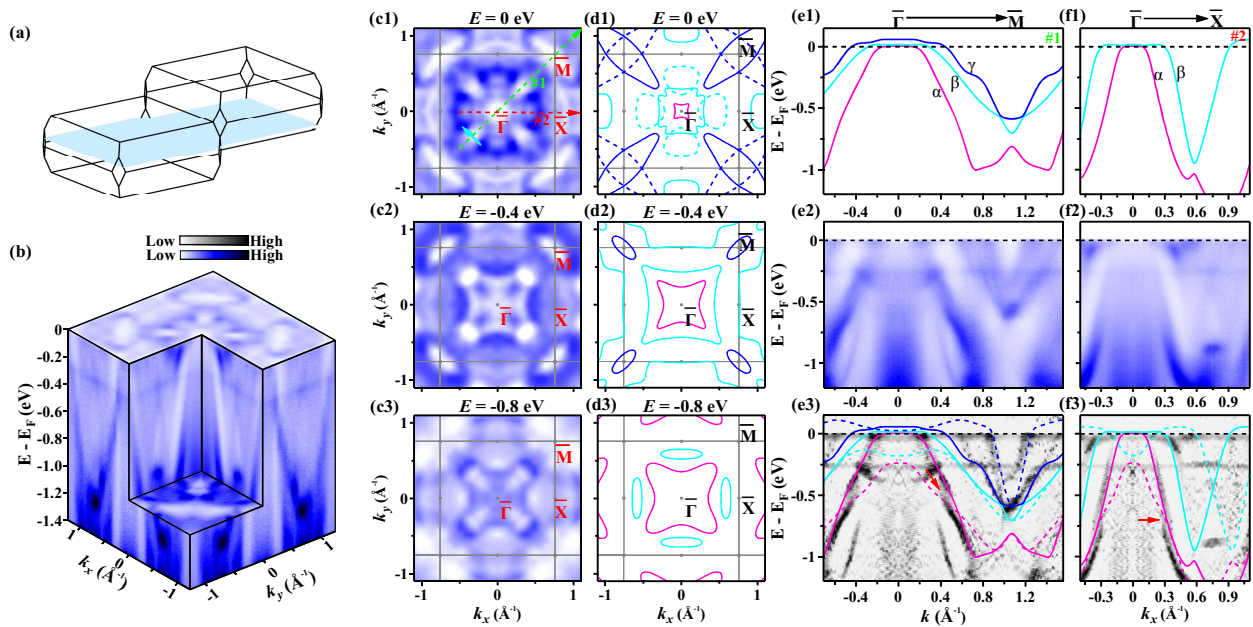

FIG. 2. (color online) (a) Illustration of the $\overline{X}$-$\overline{\Gamma}$-$\overline{X}$ plane in the bulk BZ highlighted in blue (close to Γ point). (b) 3D volume plot of the bands ($h\nu$ = 87 eV). (c1-c3) Constant energy contours of the band structure at the $E_F$, and at 0.4 and 0.8 meV below $E_F$, integrated over 10 meV. (d1-d3) The corresponding cross section of the calculated results, solid lines, in (c). (e1) and (f1) Calculated band dispersion along $\overline{\Gamma}$-$\overline{M}$ and $\overline{\Gamma}$-$\overline{X}$ directions, respectively. (e2-e3) Band structure and corresponding second derivative spectrum along the $\overline{\Gamma}$-$\overline{M}$ direction as indicated by the dashed green arrow (cut #1) in (c1). (f2-f3) Band structure and corresponding second derivative spectrum along the $\overline{\Gamma}$-$\overline{X}$ direction as by the dashed red arrow (cut #2) in (c1). The solid lines represent the original valence bands, and the dashed lines represent the nested valence bands.

The unpredicted "double" pocket feature has been observed in other materials [10, 40], which was explained in two scenarios: (i) $k_z$ broadening [40] and (ii) band folding [10]. However, the case here is unlikely to be caused by $k_z$ broadening since this effect cannot cause the hole pocket perpendicular to the $\overline{\Gamma}$-$\overline{M}$ direction near the Z plane to be projected onto the measured plane (see Fig. S2 of the Supplemental Material [30] for a detailed discussion). Here we consider the other possibility of generating "double" pockets, namely band folding. We consider the FSs folded over a commensurate wavevector $\mathbf{Q}_0$ = (0, 0, 1). This representation was chosen because previous electrical resistivity measurements suggested the formation of an AFM superzone gap below $T_{N1}$ along the $c$-direction [23]. However, the experimental temperature is higher than $T_{N1}$. To explain the inconsistency, let us consider two possibilities: (i) the magnetic moments are antiferromagnetically correlated in the paramagnetic state [41], (ii) the presence of near-surface-induced AFM occurs at temperatures above the bulk Néel temperature [42]. The dashed lines in Fig. 2(d1) represent the folded FS sheets obtained by shifting the original FS with a wavevector $\pm \mathbf{Q}_0$ = (0, 0, 1). These result in the superposition of multiple pockets, leading to a momentum-dependent FS reconstruction with clover-shaped outer hole pockets entangled with square-shaped inner hole pocket around $\overline{\Gamma}$, and a "double" pocket feature at $\overline{M}$. At the intersection of these sheets, the FS is reconstructed. Figures 2(c2) and (c3) display the FS contours for $E < E_F$. As $E$ decreases, the size of the "double" pocket decreases and disappears when $E$ is lowered by -0.8 eV, indicating electron pockets. The small $\overline{\Gamma}$-centered pocket in magenta becomes clear and more expanded at -0.8 eV. These changes are consistent with the theoretical calculation [Figs. 2(d2) and 2(d3)].

In addition to FS reconstruction, band structure reconstruction was also observed. The calculated dispersions corresponding to cut #1 and cut #2 in Fig. 2(c1) are illustrated in

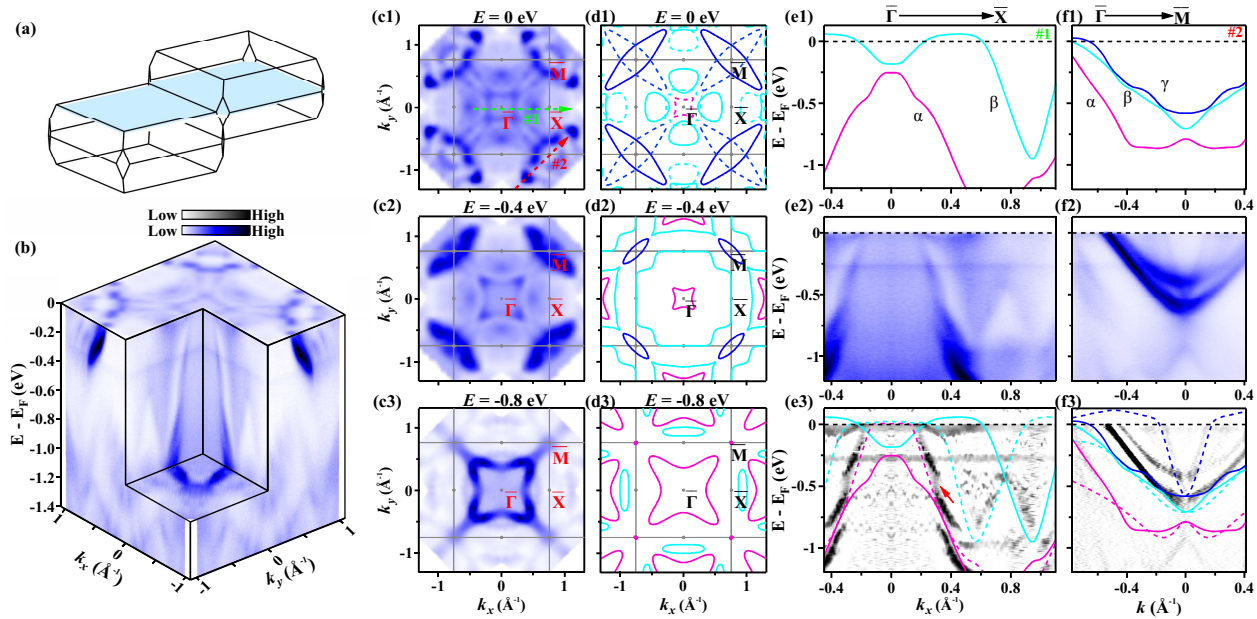

FIG. 3. (color online) (a) Illustration of the $\overline{X}$-$\overline{\Gamma}$-$\overline{X}$ plane in the bulk BZ highlighted in blue (close to $Z$ point). (b) 3D volume plot of the bands ($h\nu$ = 100 eV). (c1-c3) Constant energy contours of the band structure at the $E_F$, and at 400 and 800 meV below $E_F$, integrated over 10 meV. (d1-d3) The corresponding cross section of the calculated results in (c). (e1) and (f1) Calculated band dispersion along $\overline{\Gamma}$-$\overline{X}$ and $\overline{\Gamma}$-$\overline{M}$, respectively. (e2-e3) Band structure and corresponding second derivative spectrum along the $\overline{\Gamma}$-$\overline{X}$ direction as indicated by dashed green arrow (cut #1) in (c1). (f2-f3) Band structure and corresponding second derivative spectrum along the $\overline{\Gamma}$-$\overline{M}$ direction as indicated by dashed red arrow (cut #2) in (c1). The solid lines represent the original valence bands, and the dashed lines represent the nested valence bands.

Figs. 2(e1) and 2(f1), respectively. The two bands approaching the $\overline{\Gamma}$ point and forming hole-like pockets are indicated as $\alpha$ (magenta) and $\beta$ (cyan). The band forming an electron-like pocket around the $\overline{M}$ point is indicated as $\gamma$ (blue). Note that only the three energy bands forming the FSs are shown here. Figs. 2(e2) and 2(f2) show the detailed band structures along the two high-symmetry directions, as indicated by the dashed green (cut #1) and dashed red (cut #2) arrows in Fig. 2(c1). The corresponding second derivative spectra are shown in Figs. 2(e3) and 2(f3), respectively. To facilitate comparison with the calculation, the calculated band structure (solid lines), together with the folded bands (dashed lines), were superimposed on the second derivative spectra [Figs. 2(e3) and 2(f3)]. It can be seen that the measured and calculated band structures are in good agreement, dominated by the highly dispersive Pd 4$d$ states. The folded band (dashed blue line) surrounded the $\overline{M}$ point and formed a large electron-like pocket [Fig. 2(e3)].

Besides the band structure reconstruction, band folding can also give rise to gap opening at the intersection points between the original and folded bands. We observed the significant energy gap feature away from the $E_F$; that is, the replica bands interact with the original bands, resulting in breaks in the dispersion, as indicated by the red arrows in Figs. 2(e3) and 2(f3). The presence of the energy gap also suggests that the "double" pocket feature cannot be caused by the $k_z$ broadening. As mentioned above, some $f$-electrons are itinerant. The presence of itinerant $f$-electrons is often associated with FS instability. We consider that the imperfect FS nesting at wavevector $Q_0$ = (0, 0, 1) occurs in the overlapping region between the original and folded FSs. Since CePd$_5$Al$_2$ contains multiple Fermi pockets, we also examined the possibility of in-plane FS nesting. The in-plane FS instabilities can sometimes be identified directly by visual inspection of FSs. We observe that there are portions suitable for FS nesting. The presumed nesting vector, marked by a cyan arrow [Fig. 2(c1)], is drawn according to the neutron scattering vector $q$ = (0.235, 0.235, 0) [22, 26]. This part of the FS may play an essential role below the $T_N$. Thus, we propose that the AFM order in CePd$_5$Al$_2$ resulted from FS nesting.

To gain more information about the electronic structure and FS nesting of CePd$_5$Al$_2$, Figure 3 shows the band structure measured with a constant photon energy of 100 eV, which corresponds to a cut near the A-Z-A plane ($k_z \sim 0.9 \times 2\pi/c$) [indicated in light blue in Fig. 3(a)]. The 3D volume plot band structure in Fig. 3(b) shows the band dispersions along the high-symmetry direction. We can see that the band structures shown in Fig. 2(b) and Fig. 3(b) are significantly different, especially near the center of the BZ, suggesting the 3D character of the FSs.

The experimentally measured FSs differ significantly from the calculated FSs of the non-4$f$ reference compound LaPd$_5$Al$_2$ (Fig. 1(b)), which shows a quasi-2D character. This means that partially Ce 4$f$-electrons are involved in forming the FSs; some $f$-electrons are itinerant [29]. Figures 3(c) and 3(d) compare the measured equal constant-energy contours with calculations at different binding energies. The nested FS contour (dashed lines) at $E_F$ also added in Fig. 3(d1). The ex-perimental and computed FS contours show a good agreement regarding the shape of the pockets at $E_F$ and lower $E$.

Figures 3(e1) and 3(f1) show the calculated band structure along the high-symmetry $\overline{\Gamma}$-$\overline{X}$ (cut #1) and $\overline{\Gamma}$-$\overline{M}$ (cut #2) directions, respectively, as labeled in Fig. 3(c1). Figures 3(e2) and 3(e3) display the measured band structure and corresponding second derivative spectra along the momentum cut #1. And Figs. 3(f2) and 3(f3) present the measured band structure and corresponding second derivative spectra along momentum cut #2. Our calculated band structure (solid lines), together with the folded bands (dashed lines), have been superimposed on the second derivative spectra [Figs. 3(e3) and 3(f3)]. Along $\overline{\Gamma}$-$\overline{X}$, the measured band structure is in good agreement with the band structure calculations. While along $\overline{\Gamma}$-$\overline{M}$, there is some inconsistency between the measured and calculated band structure. A nested band (dashed cyan line) approaches the $\overline{\Gamma}$ point and forms a hole-like pocket [Fig. 3(e3)]. And a nested "V" shape band (dashed blue line) approaches the $\overline{M}$ point and forms a small electron-like pocket [Fig. 3(f3)]. Again, we observed a large energy gap due to band folding, as indicated by the red arrow [Fig. 3(e3)].

In summary, the antiferromagnetic Kondo lattice CePd$_5$Al$_2$ FSs topology and low-energy band dispersion have been investigated using high-resolution ARPES and density function theory band structure calculations. Our study reveals that the 4$f$ electrons in CePd$_5$Al$_2$ display a dual nature with both itinerant and localized characters. The obtained band dispersions and Fermi surface topology are in good agreement with DFT band calculations. ARPES measurements confirm that the Fermi surfaces, especially those in the center of the Brillouin zone, exhibit 3D characters. Our observation suggests that FS is folded over $Q_0 = (0, 0, 1)$, which is manifested by FS reconstruction, band folding and opening of the associated energy gap away from $E_F$. We propose that Fermi surface nesting drives the formation of AFM order in CePd$_5$Al$_2$. Our measurements provide key insights in understanding the nature of heavy fermion physics and its interplay with superconductivity in the antiferromagnetic superconductor CePd$_5$Al$_2$.

This work was supported by the National Natural Science Foundation of China (Grant No. 12074436), the National Key Research and Development Program of China (Grants No. 2022YFA1604200), and the Science and Technology Innovation Program of Hunan Province (2022RC3068). J. R. and P. M. O. acknowledge support through the Swedish Research Council (VR), the K. and A. Wallenberg Foundation (Grant No. 2022.0079), and the Swedish National Infrastructure for Computing (SNIC), for computing time on computer cluster Tetralith at the NSC center Linköping. Y.S. acknowledges the support from the Swedish Research Council (VR) through a Starting Grant (Dnr. 2017-05078). O.T. acknowledges support from the Swedish Research Council (VR) and the Knut and Alice Wallenberg Foundation. M.M. is partly supported by a Marie Sklodowska-Curie Action, International Career Grant through the European Commission and Swedish Research Council (VR), Grant No. INCA-2014-6426, as well as a VR neutron project grant (BIFROST, Dnr. 2016-06955). Further support was also granted by the Carl Tryggers Foundation for Scientific Research (Grants No. CTS-16:324 and No. CTS-17:325). Work at Los Alamos was performed under the auspices of the U.S. Department of Energy, Office of Basic Energy Sciences, Division of Materials Sciences and Engineering.